\documentclass[conference]{IEEEtran}
\usepackage{ifpdf}
\usepackage{multirow}

\usepackage{cite}

\ifCLASSINFOpdf
   \usepackage[pdftex]{graphicx}
\else
   \usepackage[dvips]{graphicx}
\fi
\usepackage{amsmath}
\usepackage{amsfonts}
\usepackage{amssymb}
\usepackage{array}
\usepackage{url}

\begin{document}
%
\title{MmWave MU-MIMO for Aerial Networks}

\author{\IEEEauthorblockN{Travis Cuvelier and Robert W. Heath Jr.}
	\IEEEauthorblockA{Department of Electrical and Computer Engineering\\
		The University of Texas at Austin\\
		Austin, TX 78712-1084 USA\\
		Email: tcuvelier@utexas.edu, rheath@utexas.edu}}


%


\maketitle

\begin{abstract}
	Millimeter wave offers high bandwidth for air-to-air (A2A) communication. In this paper, we evaluate the rate performance of a multiuser MIMO (MU-MIMO) configuration where several aircraft communicate with a central hub. We consider a hybrid subarray architecture, single path channels, and realistic atmospheric attenuation effects. We propose a mathematical framework for the analysis of millimeter wave (mmWave) MU-MIMO networks. Via Monte Carlo simulation, we demonstrate that mmWave is a promising technology for delivering gigabit connectivity in next-generation aerial networks.
	
\end{abstract}

%
\IEEEpeerreviewmaketitle

\section{Introduction}

High data rates are important for connecting aerial vehicles for applications like temporary mobile cellular coverage \cite{uavCellular} or cooperative sensing. Unfortunately, lower frequency A2A solutions offer limited rates due to limited available bandwidth.

In this paper, we make the case that mmWave can offer gigabits-per-second data rates in A2A networks. We start by developing an appropriate receive signal model that incorporates weather dependent atmospheric effects. Using this model, we develop a mathematical framework for evaluating rates in a hierarchical network topology where one aircraft, acting as a base station, serves other aircraft via MU-MIMO with a subarray based hybrid beamforming architecture (see, for example \cite{robertOverview}). Finally, we provide estimates of the variation of the expected per-user and network achievable rates with the number of users. 

There is limited prior work on mmWave for A2A networks, and in particular there are no well-established benchmarks. 
 Motivated by high aircraft mobility, fast beam training strategies are proposed in \cite{enablingUAVCellular}. While such insights are valuable for A2A waveform design, the numerical results provided do not provide estimates of achievable rates under realistic network configurations. Several references, e.g. \cite{satprop}, discuss the use of mmWave for earth-space communication. These sources, and some in the radar literature, e.g. \cite{radarbk}, provide an overview of atmospheric effects relevant to mmWave LOS propagation--lending valuable insight into the A2A channel. The literature on aeronautical channel modeling at lower frequencies is sparse.  While \cite{haas} considers both A2A and air-to-ground (A2G) communication while an aircraft is underway, the Rician model parameters are obtained from A2G measurements with lower frequency non-directive antennas. In \cite{rice}, an extended Rician model is applied to the UAV A2A channel at 2.4 GHz. The model accounts for the diminishing influence of the non-line-of-sight (NLOS) signal components as altitude increases. In our work, we consider higher frequencies and higher elevations such that the NLOS component may be safely neglected. 
 
There are also several relevant references that focus on A2G communication. The excellent survey \cite{guvenc} provides a comprehensive overview of A2G channel modeling topics relevant to UAVs. While much of the focus is on lower frequencies, there is some commentary on mmWave systems. Several references, e.g. \cite{dcell}, discuss the prospect of delivering cellular coverage using UAVs. Motivated by this, \cite{sga2g} and \cite{ca2g} use stochastic geometry to analyze coverage probabilities in an A2G scenario. In \cite{sga2g}, a model is analyzed where some users receive downlink service from a single UAV, while others communicate over device-to-device links. The coverage analysis is used to design mission plans for a single mobile airborne platform. Similarly, \cite{ca2g} models downlink intercell interference in a network where UAVs serve as cellular base stations. While attention has been paid to the air-to-ground aspects of UAV delivered cellular service, our work is applicable to the inherently necessary wireless backhaul \cite{ca2g}. In particular, we envision a scenario where several UAVs providing cellular coverage communicate with an aerial gateway. An aerial gateway could, in absence of local terrestrial infrastructure, provide a link to terrestrial networks via satellite. Our work is also applicable to a scenario where individual UAVs act as communication relays from ground users to an airborne command center. 
 
\section{System model}
\subsection{MIMO configuration and aerial network topology}
\begin{figure}[h]
	\centering
	\includegraphics[scale = .51]{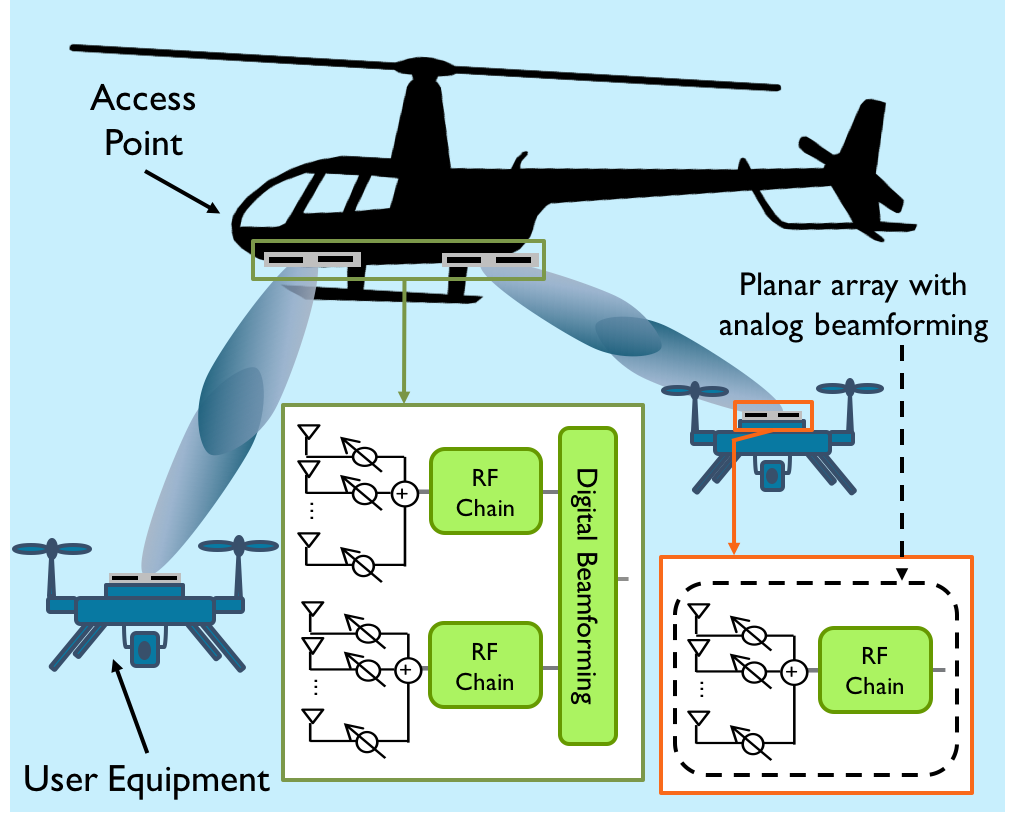}
	\caption{We assume a star network topology where an aerial access point (AP) communicates with several airborne users (UEs). We assume that each user has only one RF chain (only analog beamforming) while the more complex AP has $N_{\text{UE}}$ subarrays, with one RF chain for each. We assume these subarrays are placed in different locations onboard the aircraft, and neglect any mutual coupling between subarrays. Furthermore, we assume the UE and AP arrays are oriented with their array normals along the vertical. We focus our attention on an \textit{uplink} (UL) scenario where the communication from the aerial users to the access point.  }
	\label{fig:sys}
\end{figure}

Due to hardware constraints, mmWave MIMO systems often make use of a hybrid beamforming architecture where the number of antennas exceeds the number of RF chains \cite{robertOverview}. We assume a subarray based architecture where each RF chain is connected to a subset of antennas, with each subset equipped with analog beamforming. We assume that the AP subarrays and UE arrays themselves are uniform $\lambda/2$ spaced square planar arrays of patch radiators. While these are typical assumptions (see \cite{robertOverview}), it is straightforward to extend our approach to other array geometries. We assume that the analog beamformers can vary both phase and magnitude, relaxing non-convex constraints imposed by other architectures \cite{robertOverview}.  We denote the number of users in the network as $N_{\text{UE}}$. Figure~\ref{fig:sys} illustrates our assumptions in terms of network topology and MIMO configuration.

\subsection{Digital received signal model}\label{sec:rcvd}
At sufficient altitude, we expect that the mmWave A2A channel will consist of only LOS propagation. In LOS communication, the gain patterns of the transmit (TX) and receive (RX) antenna systems play a preeminent role in signal propagation.  Our received signal model incorporates the gain patterns of the AP subarrays and UE arrays after analog beamforming-- accounting for mutual coupling and thus more accurately characterizing TX and RX powers and SNRs \cite{circuitComms}\cite{ptxp}. Therefore, we can obtain physical results like those in \cite{circuitComms} without resorting to a circuit theoretic treatment. We ignore coupling between elements of different subarrays. We assume perfect carrier and frame synchronization among the users. This allows us to extend our single path channel model to the broadband case.  We assume that the aircraft have limited acceleration over the channel coherence interval-- for LOS communication and motion at constant velocity, the effects of Doppler are analogous to carrier frequency offset and are presumed to be mitigated by the carrier synchronization. While we do not model temporal variation, it is worth noting that the coherence time of the channel is not fatally short for large bandwidths, even in extreme cases. The LOS channel coherence time will be limited primarily by he beam alignment itself \cite{vutha}. Even with higher aircraft relative velocities ($\sim1000$ kmph), lower mmWave wavelengths ($.5$ mm), and narrow beams ($1$ degree), \cite{vutha} indicates that for reasonable TX/RX separations ($> 15$ meters)  the LOS channel coherence time will exceed half a millisecond. Finally, it is notable that while our LOS propagation assumption implicitly assumes a sparse channel, we do not explicitly exploit this property (through, for example, the development of compressive algorithms). 

We begin by deriving a digital channel model for communication between a pair of TX and RX arrays with analog beamforming. First, we separate the effective end-to-end digital channel $h$ into the large scale term $\alpha$ and a small scale term $e^{j\beta}$.
The model becomes
\begin{align}\label{eq:lss}
h = \alpha e^{j\beta},
\end{align}where $\alpha$ is dependent on the TX and RX analog beamforming and path loss. We follow by extending this to the case of a single UE communicating with an AP with multiple subarrays, and continue with an extension to MU-MIMO.

\subsubsection{Large scale attenuation}\label{lsa}

Let $G_{\text{tx}}$ and $G_{\text{rx}}$ denote the TX and RX array directivities, let $P_{\text{tx}}$ denote the TX radiated power (in Watts), and let $L(r)$ account for path losses. Assuming free space propagation, the Friis Equation gives the received power, $P_{\text{rx}}$,  as \cite{bal}\cite{pozar}
\begin{align}
P_{\text{rx}} = P_{\text{tx}}G_{\text{tx}}(\boldsymbol{\hat{\theta}_{\text{tx}}},\mathbf{f})G_{\text{rx}}(\boldsymbol{\hat{\theta}_{\text{rx}}},\mathbf{w})\left(\frac{\lambda}{4\pi}\right)^2L (r).
\end{align} Assuming lossless arrays we refer to $G_{\text{tx}}$ and $G_{\text{rx}}$ as the gains (rather than directivities) of the TX and RX arrays \cite{pozar}. The gains are functions of the  beamforming and combining vectors $\mathbf{f}$ and $\mathbf{w}$, as well as the angular bearings from transmitter to receiver, $\boldsymbol{\hat{\theta}}_{\text{tx}} = (\theta_{\text{tx}\rightarrow \text{rx}},\phi_{\text{tx}\rightarrow \text{rx}})$ and vice versa, $\boldsymbol{\hat{\theta}}_{\text{rx}} = (\theta_{\text{rx}\rightarrow \text{tx}},\phi_{\text{rx}\rightarrow \text{tx}})$. $L(r)$ accounts for losses as a function of range due to free space propagation and, in the case of mmWave, additional atmospheric absorption and scattering.  

For a $\lambda/2$ planar array in the x-y plane, we define the array steering vector $\mathbf{a(\boldsymbol{\hat{\theta}})}= \mathbf{a_{\text{x}}}(\theta,\phi)\otimes \mathbf{a_{\text{y}}}(\theta,\phi)$ \cite{hvt}. Letting $N_x$ and $N_y$ be the number of elements in the $x$ and $y$ directions, we have $\mathbf{a_{\text{y}}}(\boldsymbol{\hat{\theta}}) = [1\text{, }e^{-j\pi\sin(\theta)\sin(\phi)}\text{, ... }e^{-j\pi (N_y-1)\sin(\theta)\sin(\phi)}]^*$ and $\mathbf{a_{\text{x}}}(\boldsymbol{\hat{\theta}}) = [1\text{, }e^{-j\pi\sin(\theta)\cos(\phi)}\text{, ... }e^{-j\pi (N_x-1)\sin(\theta)\cos(\phi)}]^*$. Defining $\mathbf{A(\boldsymbol{\hat{\theta}})}= \mathbf{a(\boldsymbol{\hat{\theta}})a(\boldsymbol{\hat{\theta}})}^*$, we obtain a positive definite $\mathbf{Q} = \int_0^{2\pi}\int_0^\pi \mathbf{A(\boldsymbol{\hat{\theta}})} |F(\boldsymbol{\hat{\theta}})|^2 \sin(\theta)\, \mathrm{d}\theta\, \mathrm{d}\phi$ \cite{hvt} \cite{ptxp}. If $\mathbf{w}$ is the vector of array weights $|F(\boldsymbol{\hat{\theta}})|^2$ is the radiant intensity pattern of the radiators, the function $G$ for such an array can be written as \cite{bal}\cite{hvt}
\begin{align}
G(\boldsymbol{\hat{\theta}},\mathbf{w}) = \frac{\mathbf{w^*}\mathbf{a{(\boldsymbol{\hat{\theta}})}} \mathbf{a{(\boldsymbol{\hat{\theta}})}^*} \mathbf{w}}{\mathbf{w^*}\mathbf{Q}\mathbf{w}/4\pi}|F(\boldsymbol{\hat{\theta}})|^2.\label{eq:gainz}
\end{align}

In lower frequency systems, $L(r)= 1/r^2$, where $r$ is the range in meters. Millimeter wave radiation is further attenuated by atmospheric absorption and scattering. These losses are exponential with the range parameter $r$. At mmWave, a good model for $L(r)$ is
\begin{align}
L(r) = \dfrac{10^{-r\gamma/10}}{r^2} \label{eq:takeAnL}
\end{align} where $\gamma > 0$ is the so-called atmospheric specific attenuation in dB/m. We discuss $\gamma$ in Section~\ref{sec:atten}.

We write the large scale coefficient for a channel between TX/RX subarray pair as
\begin{align}
\alpha = \sqrt{G_{\text{rx}}(\boldsymbol{\hat{\theta}}_{\text{rx}} )G_{\text{tx}}(\boldsymbol{\hat{\theta}}_{\text{tx}}) L(r)}\left(\dfrac{\lambda}{4\pi}\right)
\end{align} where notably we have dropped the explicit dependence of the gains on beamforming weights.

\subsubsection{Small scale digital channel coefficient}
As we have assumed LOS communication with minimal scattering, our small scale fading term is limited to a phase. We assume that $\beta$ is distributed uniformly from 0 to $2\pi$. Further, we will assume that the phases between any AP subarray and UE array pair are independently and identically distributed. This assumption is reasonable, even when considering the small scale terms for the channels between two AP subarrays and a given user. Aircraft are highly mobile relative to the millimeter scale of $\lambda$, thus even slight variations in the AP's canting or vibrations could lead to very different relationships between the relevant phases. Furthermore, this assumption also removes all dependence on the placement of subarrrays onboard the AP.
\subsubsection{Uplink received signal models}
We denote the gain pattern of a UE array evaluated along the bearing from UE to AP as  $G_\text{u}(\boldsymbol{\hat{\theta}}_{\text{u}})$ and let $G_{\text{a}, k}(\boldsymbol{\hat{\theta}}_{\text{a}})$  denote the gain patterns of the AP subarrays evaluated in the direction of the airborne UE. 
For the single user (SU) uplink channel, $\mathbf{h_{\text{dig}}}$, we apply (\ref{eq:lss}) across the $N_{\text{UE}}$ AP subarrays to obtain the equivalent channel
\begin{align}
\mathbf{h_{\text{dig}}}  &=  {\begin{bmatrix}
\sqrt{G_{\text{a},1}(\boldsymbol{\hat{\theta}}_{\text{a}})}e^{j\beta_1} \\
\vdots \\
\sqrt{G_{\text{a},N_{\text{UE}}}(\boldsymbol{\hat{\theta}}_{\text{a}})}e^{j\beta_{N_{\text{UE}}}} \\
\end{bmatrix}}\sqrt{G_{\text{u}}(\boldsymbol{\hat{\theta}}_{\text{u}})L(r)}\left(\dfrac{\lambda}{4\pi}\right)\label{eq:su}.
\end{align}
We define $\mathbf{y_{\text{digital}}}$ as the received signal vector across the AP subarrays, $s$ as the transmitted signal with average transmit power $\mathbb{E}(|s|^2)$ , and $\mathbf{n} \in \mathbf{C}^{N_{\text{UE}} \times 1}$ as the noise vector. The received signal model is
\begin{align}
\mathbf{y}_{\text{dig}} =  \mathbf{h_{\text{dig}}}s + \mathbf{n}.
\end{align}  We assume IID circularly symmetric complex Gaussian noise $\mathbf{n} \sim \mathcal{N}(\mathbf{0}, BN_o\mathbf{I}  )$, where $N_o$ is the received noise power spectral density in Watts/Hz, and $B$ is the system bandwidth. The uncorrelatedness of the  noise follows from the assumption that the subarrays are uncoupled \cite{circuitComms}. We assume $N_o$ to be  independent of the beamforming weights at each subarray i.e. we assume a (spatially) isotropic sky noise environment \cite{circuitComms}. The calculation of $N_o$ is discussed in Section~\ref{sec:hughes}.

We extend the SU model to multiple UEs, each with a single analog beamformed array, communicating with an AP equipped with $N_{\text{UE}}$ identical subarrays. Let the k$^{\text{th}}$ element of the vector $\mathbf{s}$ be the symbol transmitted by user k over $\mathbf{h_{\text{dig, k}}}\in\mathbb{C}^{N_{\text{UE}}\times1} $, the effective uplink digital channel from user k to the AP. The $\mathbf{h_{\text{dig,k}}}$ are defined as in (\ref{eq:su}) but with bearings evaluated along the path from the AP to user k and vice versa. We define the matrix  $\mathbf{H_{\text{UL}}}$ such that its k$^{\text{th}}$ column is given by $\mathbf{h_{\text{dig,k}}}$. The MU-MIMO received digital signal $\mathbf{y}_{\text{dig}}$ is given by: 
\begin{align}
\mathbf{y}_{\text{dig}} = \mathbf{H_{\text{UL}}}\mathbf{s} + \mathbf{n}\label{eq:digmodel}
\end{align} where now each subarray receives a superposition of signals from multiple users.  

\subsubsection{Uplink beamforming and combining}\label{sec:ulbf}

Our analog beamforming strategy can be motivated by explicitly demonstrating the role played by the beamforming/combining in (\ref{eq:digmodel}). Let $\mathbf{w}_{k}$ be the combining weights for the $k^{\text{th}}$ AP subarray, and $\mathbf{f}_{k}$ be the beamforming weights for the array onboard the $k^{\text{th}}$ UE. Then, using (\ref{eq:gainz}) and (\ref{eq:digmodel}) the elements of $\mathbf{H_{\text{UL}}}$ are
\begin{multline}
[\mathbf{H_{\text{UL}}}]_{i,k} = \sqrt{\frac{\mathbf{w}_{i}^* \mathbf{A}_{a}(\boldsymbol{\hat{\theta}}_{a,k})\mathbf{w}_{i}}{\mathbf{w}_{i}^* \mathbf{Q}_{\text{a}}\mathbf{w}_{i}}\frac{\mathbf{f}_{k}^* \mathbf{A}_{\text{u}}(\boldsymbol{\hat{\theta}}_{\text{u},k})\mathbf{f}_{k}}{\mathbf{f}_{k}^* \mathbf{Q_{\text{u}}}\mathbf{f}_{k}}} \\ \times \sqrt{L(r_k)}\lambda |F(\boldsymbol{\hat{\theta}}_{\text{a},k})F(\boldsymbol{\hat{\theta}}_{\text{u},k})| e^{j\beta_{i,k}}. \label{eq:ULH}
\end{multline} 
where $L$, $F$, $\mathbf{A}_{\text{a}}, \mathbf{Q}_{\text{a}}, \mathbf{A}_{\text{u}},$ and $\mathbf{Q}_{\text{u}}$ are as defined in Sec.~\ref{lsa} for the geometries defined for the AP subarrays and UE arrays. Analogously, we denote the bearing from the $k^{th}$ UE to the AP as $\boldsymbol{\hat{\theta}}_{\text{u},k}$ and the bearing from the AP to the $k^{th}$ UE as $\boldsymbol{\hat{\theta}}_{\text{a},k}$. Again, the $\beta_{i,k}$ are independent, uniformly random phases. 

We propose a heuristic, but effective beamforming strategy for the proposed aerial network architecture. We assume that the $\boldsymbol{\hat{\theta}}_{\text{a},k}$ are known at the AP and the $\boldsymbol{\hat{\theta}}_{\text{u},k}$ are known at their respective UEs. This corresponds to side information on the relative position of the aerial UEs to the AP. This information could be obtained via coordinated control and mission planning, for example, a group of UAVs remote controlled over a common data link could take into consideration the orders relayed to other users in the network. Further, we assume $\mathbf{H}_{\text{UL}}$ is known at the AP. 

Now we describe the analog beamforming. Assuming a constraint on radiated power, i.e. $\mathbb{E}(\mathbf{s}\mathbf{s}^*) = \mathbf{I} P_{\text{tx}}$, we employ a greedy strategy at the UE. The UE's transmitter selects its beamforming weights to maximize its directivity in the direction of the access point. That is to say for each UE, the transmit precoding weights are
\begin{align}
\mathbf{f}_{k} = \arg\max_{\mathbf{f}} \frac{\mathbf{f}^* \mathbf{A}_u(\boldsymbol{\hat{\theta}}_{\text{u},k})\mathbf{f}}{\mathbf{f}^* \mathbf{Q}_{\text{u}}\mathbf{f}}.\label{eq:ueanbf}
\end{align} The maximum is attained when $\mathbf{f}_{k}$ is chosen along the eigenvector with the largest eigenvalue of $\mathbf{Q}_{\text{u}}^{-1}\mathbf{A}_u({\boldsymbol{\hat{\theta}}}_{\text{u},k})$ \cite{rayray}. 
We associate each AP subarray with a particular UE, and proceed in a greedy fashion by maximizing the directivity of the $k^{\text{th}}$ AP subarray in the direction of the $k^{\text{th}}$ user. Formally, we choose $\mathbf{w}_{k}$ to lie along the eigenvector of $\mathbf{Q}_{\text{a}}^{-1}\mathbf{A}_a({\boldsymbol{\hat{\theta}}}_{\text{a},k})$ with the largest eigenvalue.  Our strategy thus increases the magnitude of the diagonal terms in our channel matrix.

We assume linear digital combining with $\mathbf{W}^* \in \mathbb{C}^{N_{ue } \times N_{ue}}$ at the AP. The receiver processes $\mathbf{y}_{\text{digital}}$ with $\mathbf{W}^*$ and obtains
\begin{align}
\mathbf{y}_{\text{BF}} = \mathbf{W}^*\mathbf{y}_{\text{dig}} =  \mathbf{W}^*(\mathbf{H_{\text{UL}}}\mathbf{s} + \mathbf{n})\label{eq:digbfex}.
\end{align} For the digital combining, we consider two approaches. In the first approach, the combiner applies the classical MMSE beamformer in (\ref{eq:digbfex}). Defining $\text{SNR}=P_{\text{tx}}/(N_oB)$, the MMSE digital beamformer can be expressed as \cite{mmse}
\begin{align}
\mathbf{W}_{\text{MMSE}}^*= (\mathbf{I}/\text{SNR} +\mathbf{H}^*\mathbf{H})^{-1}\mathbf{H}^*.
\end{align} Given the model (\ref{eq:digmodel}), this beamformer is optimal (when coupled with successive interference cancellation) from an information theoretic (rate maximization) perspective \cite{mmse2}. We also consider the case of no digital beamforming, i.e. the application of $\mathbf{W}^*$ equal to identity.  In this case, interference is mitigated  by the analog beamforming only. Since the approach bears similarity to spatial division multiple access (SDMA), we refer to this as the ``SDMA approach".

\section{Channel model parameters }
In this section, we discuss the specific attenuation, $\gamma$, and the noise power spectral density, $N_o$. Atmospheric gasses, including diatomic oxygen and water vapor absorb mmWave radiation \cite{satprop}\cite{radarbk}. Clouds, rain, and other precipitation further absorb and scatter millimeter wave radiation \cite{satprop} \cite{radarbk}. These effects cause the received signal power to fall exponentially with increasing range--accounted for by $\gamma$  in (\ref{eq:takeAnL}). Detailed balance requires that bodies that absorb energy must also radiate it. Thus, we pay a double penalty when dealing with the aforementioned atmospheric attenuation processes--the molecules which absorb our signal power further emit incoherent noise that is picked up at a receiver.  

\subsection{Atmospheric specific attenuation}\label{sec:atten}

In this section, we assess the impact of three prominent contributors to mmWave atmospheric attenuation: namely gaseous molecular absorption, absorption and scattering from suspended liquid water in clouds/fog, and scattering from rain and other precipitation. These losses are decomposed as
\begin{align}
\gamma = \gamma_{\text{gases}} + \gamma_{\text{fog}} + \gamma_{\text{precipitation}}.
\end{align}
\subsubsection{Atmospheric gases}
The ITU has recent recommendations on computing the specific attenuations of atmospheric gases, $\gamma_{\text{gases}}$, at frequencies up to 1000 GHz \cite{ITUAT}. The models account for absorption by diatomic oxygen and nitrogen as well as water vapor. They are parameterized in terms of the partial pressures of water vapor and dry air (that is to say, the partial pressure of gases other than water vapor) as well as ambient temperature. The attenuation is quite modest at lower frequencies, however becomes quite substantial (around 1 dB/km) near 50 GHz (15 dB/km) due to an oxygen absorption line at 60 GHz. While the attenuation decreases above the 60 GHz resonance, water vapor begins to dominate due to an absorption line at 118 GHz. One can use the models in \cite{ITUAT} with the standardized reference atmospheres in \cite{ITUAtm} to gain insight into how atmospheric absorption varies with altitude. In general, there is less attenuation at higher altitudes--in part due to the decreasing density of the absorptive gasses.

\subsection{Suspended liquid water: clouds and fog}

Suspended liquid water (either in clouds or fog) scatters mmWave radiation in the Rayleigh regime \cite{radarbk} \cite{fog}. A simple empirical model for fog attenuation valid from 30 to 100 GHz and temperatures between -8 $^{\circ}$C and 25 $^{\circ}$ is  given in \cite{fog} and \cite{fog2}. At frequencies above 100 GHz, attenuation from water vapor dominates, and at lower frequencies the attenuation is not appreciable \cite{fog}. The likelihood of suspended liquid water at temperatures outside the range is also small \cite{fog}. 


\subsection{Precipitation}
Rain attenuation varies with rain rate, velocity, drop size, canting, and shape \cite{radarbk}. The ITU makes recommendations in \cite{ITU2} and \cite{iturain} relevant to rain attenuation. Rain attenuates horizontally (with respect to the ground) polarized signals more than vertical ones \cite{radarbk}. Rain attenuation generally increases with frequency. It becomes more significant with heavy rain rates (above 10 mmph)--e.g. at 50 mmph it reaches 1 dB/km at 10 GHz. For equivalent (melted) precipitation rates, attenuation due to frozen precipitation is substantially less than that caused by equivalent rainfall, becoming comparable at higher frequencies \cite{rand}.

\subsection{Noise}\label{sec:hughes}

Noise in wireless receivers originates from sources external to the antenna array and from the receiver itself. These effects are captured by the antenna array noise temperature $T_{\text{ant}}$ (in Kelvin) and receiver noise figure $\eta$ (in dB) in the definition of the noise power spectral density, e.g. \cite{pozar}
\begin{align}
N_o = k(T_{\text{ant}}+ T_o(10^{\frac{\eta}{10}}-1))
\end{align} where $k$ the Boltzmann constant in $\text{Watts}/(\text{Kelvin}\times \text{Hz})$, and $T_o = 290$ K. In \cite{circuits}, a receiver frontend with a 7 GHz bandwidth and a $\eta=7.1$ dB noise figure is designed and prototyped. Taking inspiration from this, we assume that $T_o(10^{\frac{\eta}{10}}-1) = 1197.3$ \text{K}.

In general, $T_{\text{ant}}$ is a complicated function of environmental conditions and the beamforming weights. In general it increases with $\gamma$ \cite{satprop}. In this work, we make the simplifying assumption that $T_{\text{ant}}=T_{\text{mr}}$,  the atmospheric mean radiating temperature. Thus, we assume a spatially isotropic noise environment, allowing validation with \cite{circuitComms}. The assumption should be pessimistic for skyward directed bearings sufficiently away from the sun or moon, quite accurate for beams directed along the horizon, and reasonable for beams directed towards Earth's surface \cite{itunoise}. For clear and cloudy weather, $T_{\text{mr}}$ can be estimated as a linear function of the surface temperature \cite{itunoise}. For typical surface conditions, these models give $T_{\text{mr}}$ around 300 K, implying that the antenna noise contributes around 20$\%$ of the system noise. For rainy atmospheres, it is reasonable to assume a $T_{\text{mr}}$ between $270$ and $280$ K \cite{itunoise}.

\section{Analytical framework}

\begin{figure}[h]
	\centering
	\includegraphics[scale = .40]{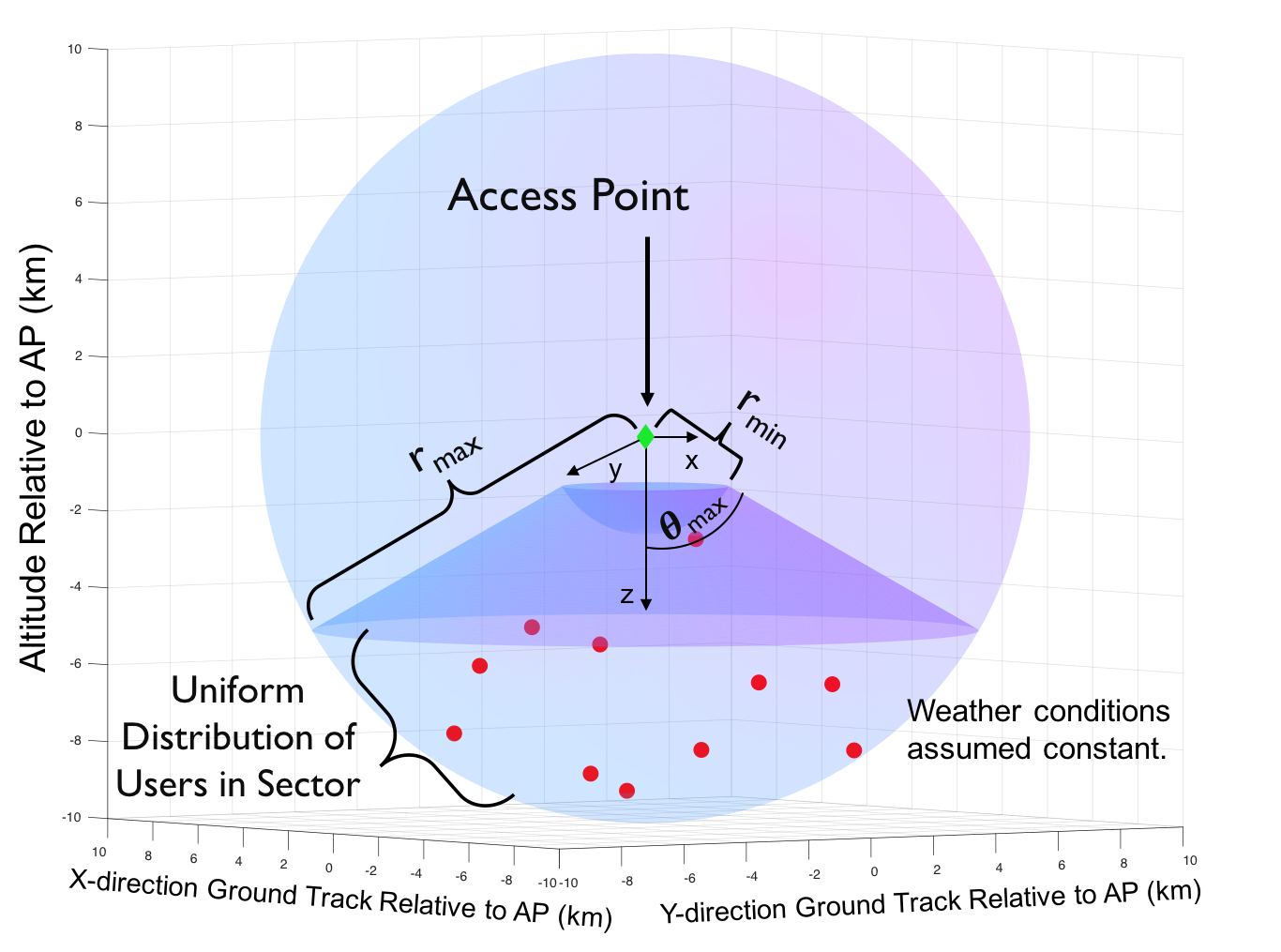}
    \vspace{-.3cm}
	\caption{Our geometric frameworks considers UEs (red dots) independently uniformly distributed in a spherical cone below the AP (green diamond). }
	\label{fig:geo}
\end{figure}

With our physical model in hand, we propose a stochastic geometry based framework for the analysis of mmWave A2A networks. We place the AP at the origin, and take $\hat{z}$ to be the direction of decreasing altitude. Antenna arrays at the access point are assumed to be oriented along the $x-y$ plane with a normal in the $\hat{z}$ direction. An illustration  is shown in Figure~\ref{fig:geo}.

We assume that there are a fixed number of UEs located below the AP in a spherical shell defined via $\theta < \theta_{\max}$ and $ r_{\min} \le r \le r_{\max}$. We take a UE's position to be uniformly randomly distributed within the shell. Requiring that the probability of a UE lying within a given $\delta r$,  $\delta\theta$, $\delta\phi$ be proportional to the differential volume swept out by those coordinates leads us to the joint distribution $f_{r,\theta,\phi}(r, \theta, \phi )	= r^2\sin(\theta)/V_{\text{shell}}$
for $r_{\min} \le r \le r_{\max}$ and $0\le \theta \le \theta_{\max}$.
Marginalizing, we obtain distributions for a user's coordinates $r$, $\theta$, and $\phi$ as
\begin{align}
f_r(r)	= 3\frac{r^2}{r_{\max}^3-r_{\min}^3} \text{, for $r_{\min} \le r \le r_{\max}$ }\label{eq:fr},
\end{align} 
\begin{align}
f_\theta(\theta )	= \frac{\sin(\theta)}{1-\cos(\theta_{\max})}\text{, for $0\le \theta \le \theta_{\max}$},	
\end{align} 
\begin{align}
f_\phi( \phi )	= \frac{1}{2\pi} \text{, for $0 \le \phi \le 2\pi$}.
\end{align} Since  $f_r(r)f_\theta(\theta )f_\phi( \phi ) = f_{r,\theta,\phi}(r, \theta, \phi )$, the coordinates are independent. UEs are assumed to have their antenna arrays located along the $x-y$ plane, directed along $-\hat{z}$. We assume that the atmospheric conditions are constant within the shell.

Stochastic geometry can be used to give insight into achievable rates. In what follows we apply our framework to a sensible scenario and, dispensing with formal analysis, estimate rates through Monte Carlo experimentation.  

\section{Simulations}\label{sec:ac1}
Carrying out the linear digital beamforming in (\ref{eq:digbfex}) leads to $N_{\text{UE}}$  channels. When combining with $\mathbf{W}^* = \mathbf{W}^*_{\text{MMSE}}$, the channels have signal-to-interference-plus-noise ratios (SINR) of
\begin{align}
\rho_{\text{MMSE},k} = \frac{\text{SNR}}{[(\mathbf{H}^*\mathbf{H}+\mathbf{I}/\text{SNR})^{-1}]_{k,k}}-1,
\end{align} where $k$ ranges from $1$ to $N_{\text{UE}}$ \cite{mmse}. Conversely, upon applying $\mathbf{W}^*$ equal to identity in (\ref{eq:digbfex}), the SINRs are
\begin{align}
\rho_{\text{SDMA},k} = \frac{P_{\text{tx}}|[\mathbf{H}]_{k,k}|^2 }{N_oB+P_{\text{tx}}\sum_{j\ne k}|[\mathbf{H}]_{k,j}|^2},
\end{align} where again $k$ ranges from $1$ to $N_{\text{UE}}$.  Assuming Gaussian signaling and $N_{\text{UE}}$ independent Gaussian decoders, the maximum per-user achievable rates for MMSE beamforming are
\begin{align}
C_{\text{MMSE},u} = B\log_2(1+\rho_{\text{MMSE},u}).
\end{align}  Summing $C_{\text{MMSE},u}$ over the UEs gives the network sum rate. The per-user achievable rates for our SDMA approach, $C_{\text{SDMA},u}$, and the corresponding sum rates are defined analogously.

\begin{figure}[h]
	\centering
	\includegraphics[scale = .157]{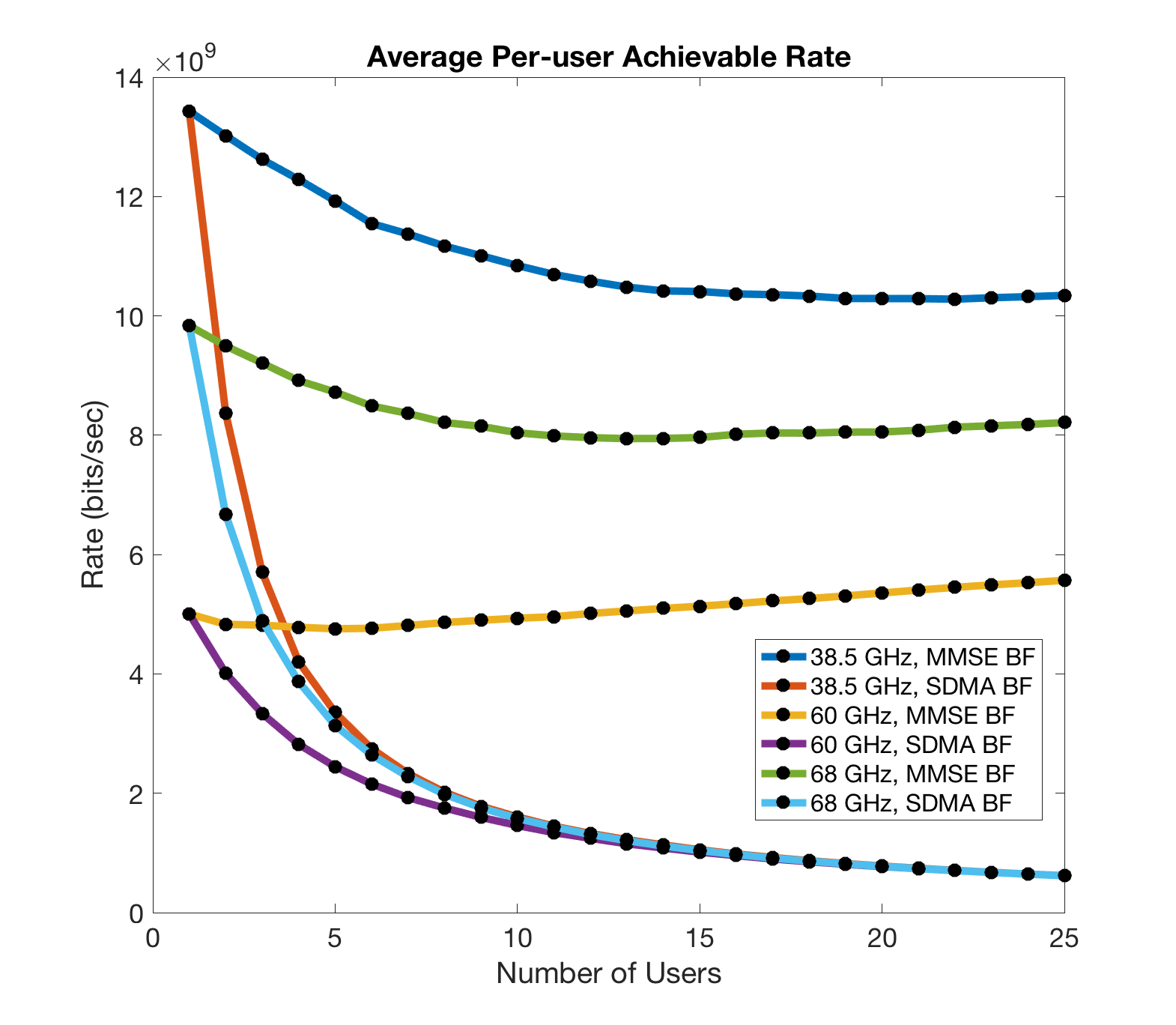}
	\vspace{-.3cm}
	\caption{This plot shows the average rate available to a single user as more users are added to the network. The rate available to a user when MMSE beamforming is used actually increases (easiest to perceive at 60 GHz) as the number of users in the network increase. For every UE added to the network, we added a $4\times 4$ subarray to the access point-- when noise limited, the increased antenna diversity compensates for the increase in interference. Without the digital beamforming (i.e. the SDMA approach), we get more intuitive behavior-- adding users increases interference and reduces the per-user achievable rate.  }
	\label{fig:res1}
\end{figure}

\begin{figure}[h]
	\centering
	\includegraphics[scale = .157]{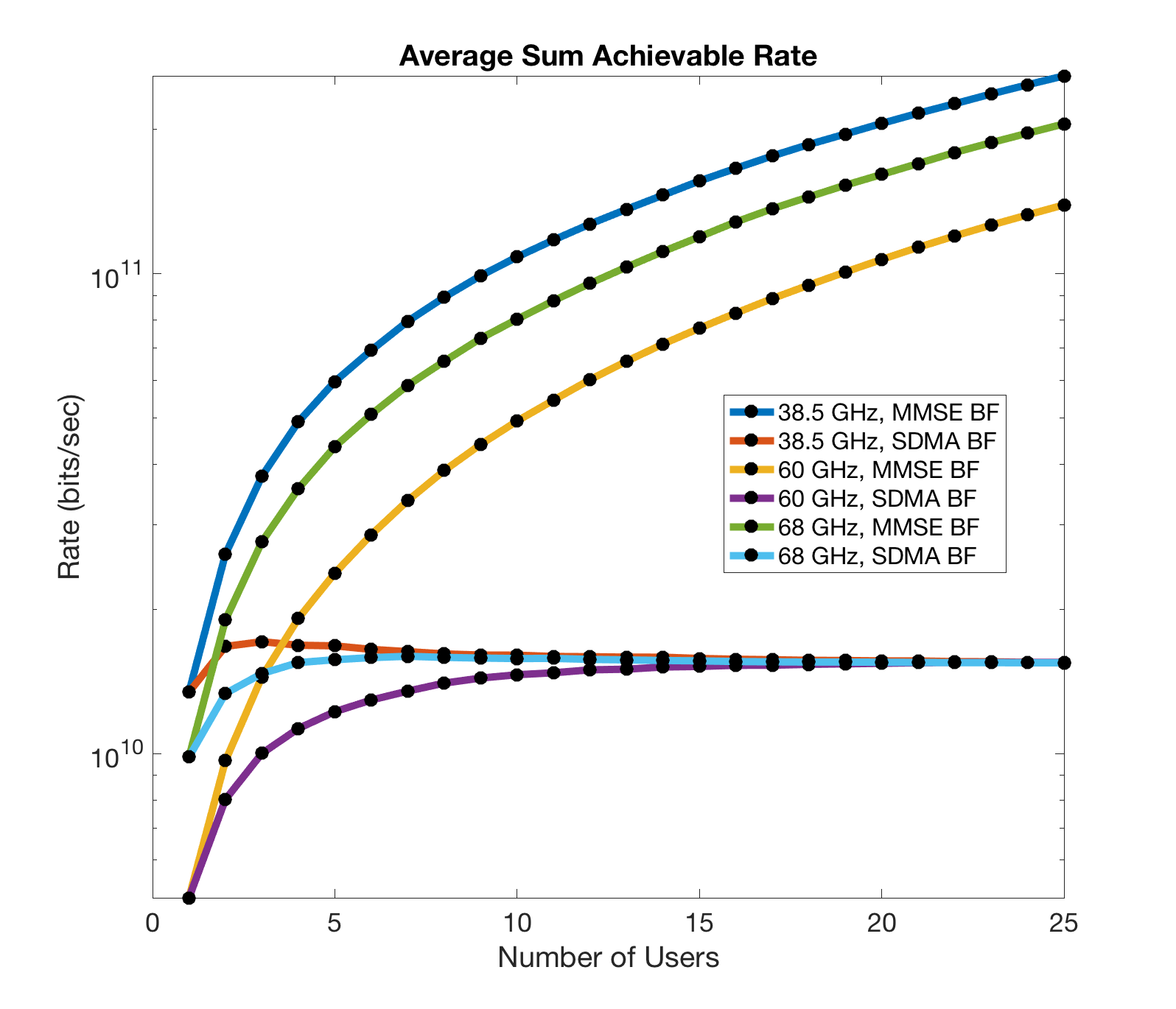}
    \vspace{-.3cm}
	\caption{This plot shows the average sum achievable network rate as a function of the number of users in the network. A diminishing return is notable, although much more so for the SDMA approach. }
	\label{fig:res2}
\end{figure}

We simulated an aerial network with a varying number of users under the assumptions discussed in the previous section and computed the set of $C_{\text{MMSE},u}$ and  $C_{\text{SDMA},u}$ for each realization. The simulation parameters are shown in Table~\ref{tab:sims}. We used antenna element patterns (i.e. $F(\boldsymbol{\hat{\theta}})$ in $(\ref{eq:gainz})$) obtained from the patch antenna design procedure in \cite{bal}.  We averaged over 1000 trials to obtain the results shown in Figs.~\ref{fig:res1} and~\ref{fig:res2}. 

Our simulations demonstrate that mmWave is a promising technology for delivering gigabit A2A connectivity. Our atmospheric parameters are typical of fair weather mid-latitudes near sea level, and our MIMO and network configurations are reasonable for networks of small, low complexity users over spread over a relatively small area. Assuming that an AP can support one subarray for each user, our results indicate that our network is not strongly interference limited when MMSE beamforming is applied. This indicates that mmWave shows promise as a communication catalyst for aerial swarms.

This analytical framework can be used to simulate other beamforming strategies and mission plans. We made a strong assumption in assuming infinite magnitude and phase resolution in our analog beamformers, and it is relatively straightforward to extend our analysis to the case where we are restricted to a codebook of beams. Our framework is also useful in analyzing different mission plans: the achievable rates will depend on both the radius and angular extent of the shell served by the access point. With a restricted codebook of beams, these effects will be more pronounced.

\begin{table}[h]
	\begin{center}
\begin{tabular}{ |lll| }
	\hline
	 \multirow{3}{*}{General} & Number of Trials & 1000 \\
	& TX Power   & 10 W\\
	& Bandwidth & 2 GHz\\ \hline
	\multirow{3}{*}{Geometry} & $r_{\min}$ & 0 km \\
	& $r_{\max}$  & 1 km \\
	& $\theta_{\max}$ & 30$^{\circ}$ \\ \hline
	\multirow{5}{*}{Weather} & Barometric Pressure & 1 atm \\
	& Ambient Temperature & 295 K \\ & Relative Humidity & 50$\%$ \\
	& $T_{\text{mr}}$ & 276 K via \cite{itunoise} \\ 
	& Rain/Fog & None \\\hline 
	
	\multirow{3}{*}{Spec. Atten.} & 38.5 GHz & .15 dB/km \\
	& 60 GHz & 14 dB/km \\ & 68 GHz & .87 dB/km \\ \hline
	\multirow{3}{*}{Antennas} & UE MIMO& 1 2$\times$2 array/UE\\
	& AP MIMO & $N_{\text{UE}}$ 4$\times$4 arrays \\ & Radiating Elements & Patch antennas \cite{bal}  \\
	\hline
\end{tabular}
\end{center}
\caption{Simulation Parameters} 
\label{tab:sims} 
 \vspace{-.5cm}
\end{table}

\section{Acknowledgments}
This work was supported in part by the Lockheed Martin Corporation, the National Science Foundation under Grant No. CNS-1731658, and an Engineering Doctoral Fellowship from the University of Texas at Austin Cockrell School of Engineering. 

\bibliographystyle{IEEEtran}
\bibliography{propSources}

\end{document}